\documentclass{cjaa}                   

\usepackage{graphicx}                   
\input{epsf.sty}                        
\input{psfig.sty}                       
\newcommand{\Msolar}{\mbox{\,$\rm M_{\odot}$}}        
\newcommand{\kms}{km~s$^{-1}$}

\begin{document}
\title{SDSS J143030.22-001115.1: A misclassified narrow-line Seyfert 1 galaxy with flat X-ray spectrum}
\author{Wei-Hao Bian, Quan-Ling Cui \& Li-Hua Chao}
\institute{Department of Physics and Institute of Theoretical
Physics, Nanjing Normal University, Nanjing 210097, China}

\date{Received ...; accepted ...}
\titlerunning{A misclassified NLS1: SDSS J143030.22-001115.1}
\authorrunning{Bian, Cui \& Chao }

\abstract{We used multi-component profiles to model H$\beta$ and
[O III]$\lambda \lambda $4959,5007 lines for SDSS
J143030.22-001115.1, a narrow-line Seyfert 1 galaxy (NLS1) in a
sample of 150 NLS1s candidates selected from the Sloan Digital Sky
Survey (SDSS) Early Data Release (EDR). After subtracting the
H$\beta$ contribution from narrow line regions (NLRs), we found
that its full width half maximum (FWHM) of broad H$\beta$ line is
nearly 2900 \kms, significantly larger than the customarily
adopted criterion of 2000 \kms. With its weak Fe II multiples, we
think that SDSS J143030.22-001115.1 can't be classified as a
genuine NLS1. When we calculate the virial black hole masses of
NLS1s, we should use the H$\beta$ linewidth after subtracting the
H$\beta$ contribution from NLRs. \keywords{galaxies:
active---galaxies: emission lines---galaxies: nuclei--- galaxies:
Seyfert---galaxies:individual: J143030.22-001115.1} } \maketitle

\section{Introduction}

Narrow-line Seyfert 1 galaxies (NLS1s) were defined by their
optical spectral characteristics: H$\beta$ full width half
maximum(FWHM) less than 2000 \kms; strong optical FeII multiples;
the line-intensity ratio of [O III]$\lambda 5007$ to H$\beta$ less
than 3 (Osterbrock \& Pogge 1985; Goodrich 1998). These properties
suggested that NLS1s likely contain less massive black holes with
higher Eddington ratios (Pounds et al. 1995; Wandel \& Boller
(1998); Laor et al. 1997; Mineshige et al. 2000), putting NLS1s at
one extreme end of the so-called Boroson \& Green (1992)
eigenvector 1. Their locus in the $M-\sigma$ relation plane is
possible special relative to the other AGNs(e.g. Bian \& Zhao
2004; Grupe \& Mathur 2004; Greene \& Ho 2005c). X-ray observation
of NLS1s revealed a strong X-ray excess and NLS1s generally have
softer X-ray spectra than other active galactic nuclei (AGNs)
(Boller et al. 1996; Leighly 1999). Grupe et al. (2004) found that
ultra-soft X-ray selection method can be used to find large
numbers of NLS1s.

Williams et al. (2003) presented a sample of 150 NLS1s candidates
found within Sloan Digital Sky Survey (SDSS) Early Data Release
(EDR; Stoughton et al. 2002), which is the largest published
sample of NLS1s. 17 of these SDSS NLS1s are observed with Chandra
and it is suggested ultrasoft X-ray emission is not a universal
characteristic of NLS1s (Williams et al. 2004). For two objects
with the smallest photon indices, SDSS J125943.59+010255.1 and
SDSS J143030.22-001115.1, the photon indices are less than one.
For this sample of 150 SDSS NLS1s, we have used multi-component
profiles model to carefully analysis the narrow line regions
(NLRs) outflow relative to broad line regions (BLRs) (Bian, Yuan
\& Zhao 2005). For SDSS J143030.22-001115.1, the line-intensity
ratio of [O III]$\lambda$5007 to narrow H$\beta$ is the largest,
up to 9.6.

In this paper, we presented our multi-component profiles modelling
for SDSS J143030.22-001115.1. The spectral analysis method is
introduced in Sect. 2. The results of profile fitting are given in
Sect. 3. Our discussion is presented in the last section. All of
the cosmological calculations in this paper assume $H_{0}=75 \rm
{~km ~s^ {-1}~Mpc^{-1}}$, $\Omega_{M}=0.3$, $\Omega_{\Lambda} =
0.7$.

\section{DATA and ANALYSIS}
Using the SDSS Query Tool, Williams et al. (2003) selected objects
fron SDSS EDR that were flagged as QSOs and that showed narrow
H$\beta$ lines. They then directly measured the width of the line
halfway between the fitted continuum and the H$\beta$ line peak as
the FWHM(H$\beta$). At last, they obtained a sample of 150 SDSS
NLS1s. SDSS J143030.22-001115.1 is one object in this sample and
its spectrum is obtained from SDSS Data Release 3 (DR3; Abazajian
et al. 2005) (See left panel in Fig. 1).

There generally exists strong Fe II multiples in the NLS1s opticla
spectra. And at the same time, there exists the asymmetry of [O
III] and/or H$\beta$ lines. Therefore, we reduced its spectrum of
SDSS J143030.22-001115.1 by the multi-component fitting task
SPECFIT (Kriss 1994) in the IRAF-STS package. The used components
are: (1) the Galactic interstellar reddening curve; (2) Fe II
template; (3) power-law continuum; (4) three sets of two-gaussian
profiles for [O III]$\lambda\lambda$4959, 5007 and H$\beta$ lines.
We take the same linewidth for each component, and fix the flux
ratio of [O III]$\lambda$4959 to [O III]$\lambda$5007 to be 1:3.
Firstly we didn't consider the starlight contribution because of
no obvious stellar lines (Gu et al. 2005). For more detail, please
referee to Bian, Yuan \& Zhao (2005; 2006).

\begin{figure}
\begin{center}
\includegraphics[width=5cm,angle=-90]{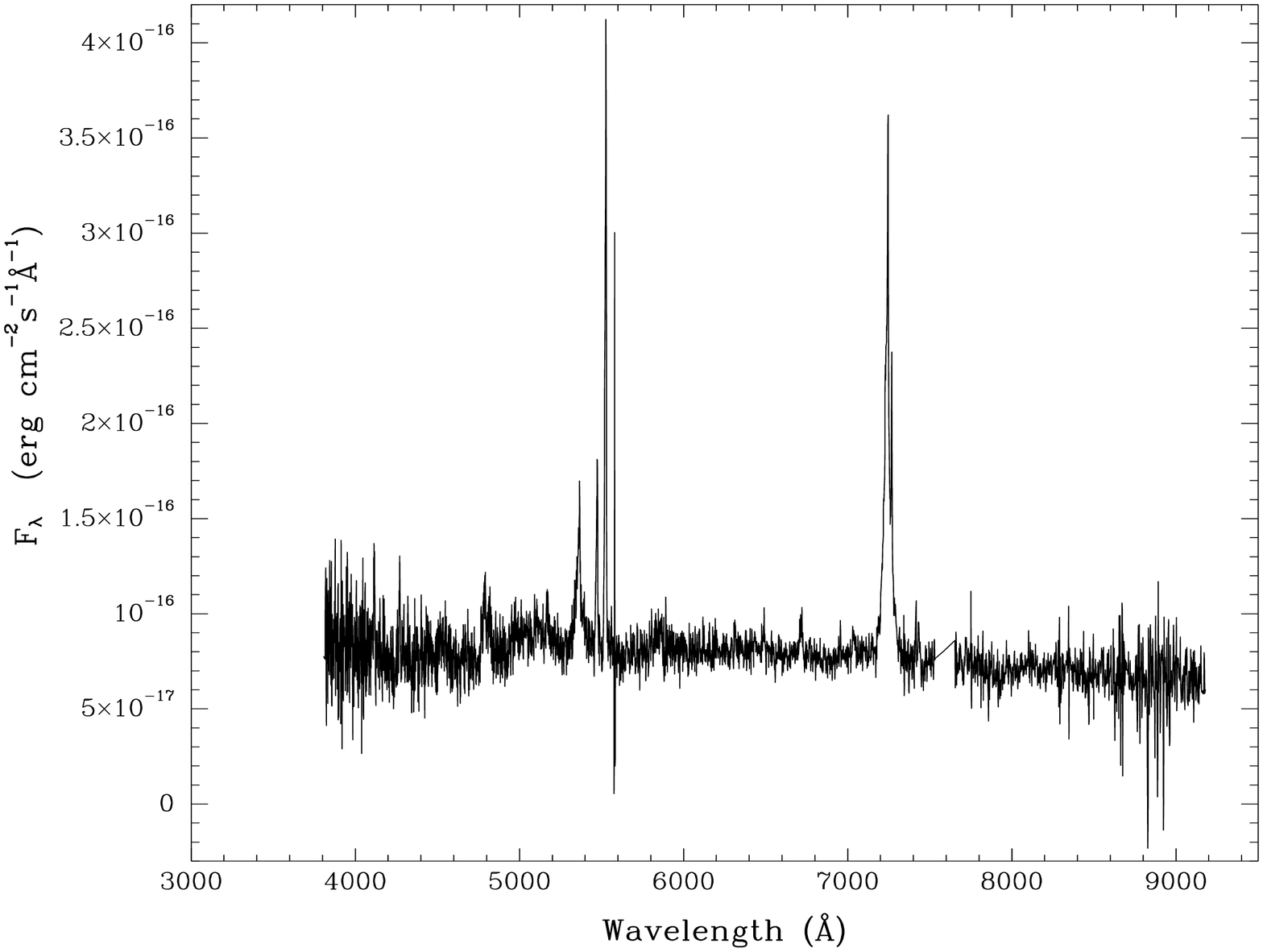}
\includegraphics[width=5cm,angle=-90]{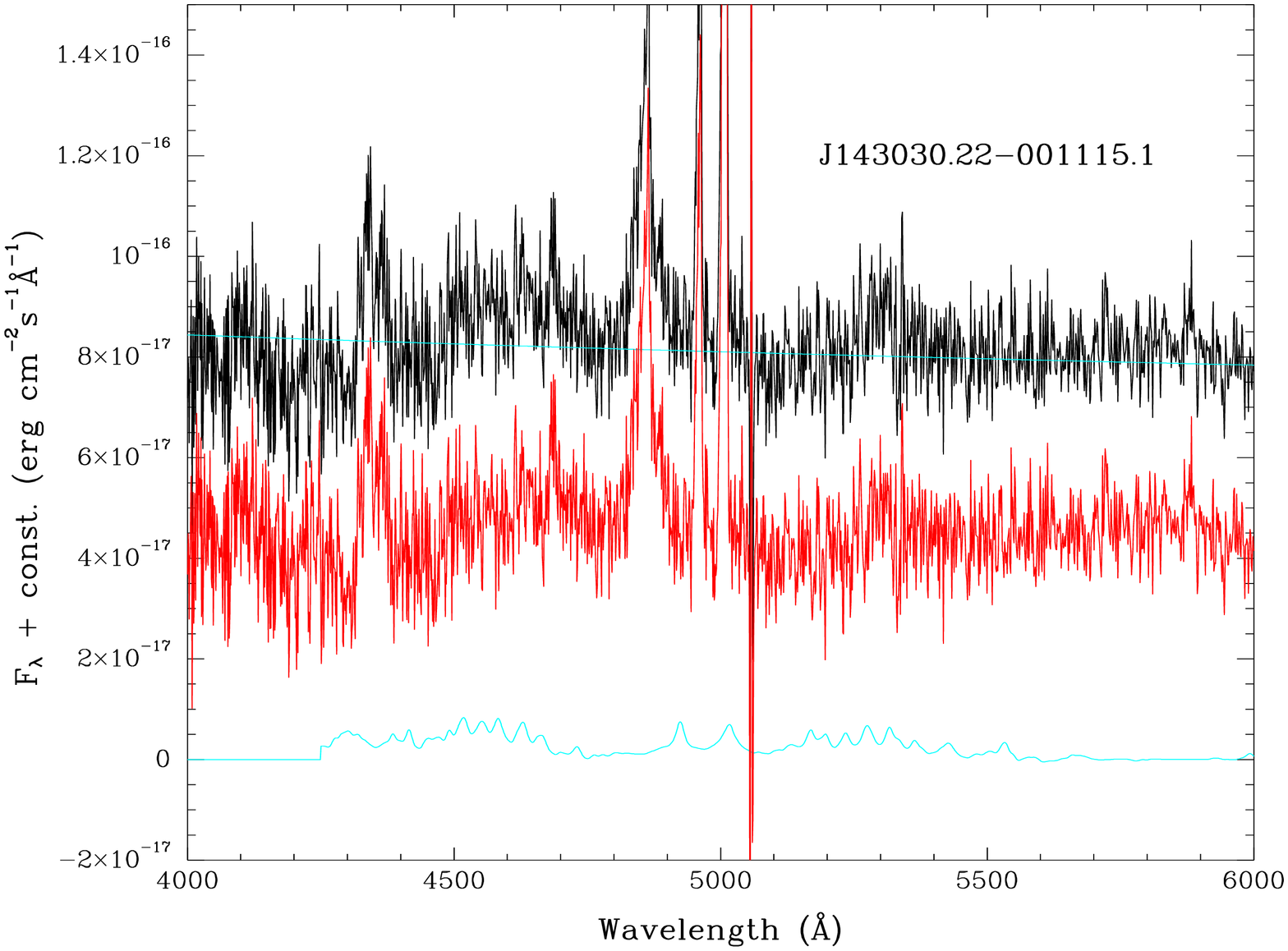}
\label{fig1}

\caption{Unprocessed spectrum of SDSS J143030.22-001115.1 (left).
Rest-frame spectra of SDSS J143030.22-001115.1 (right): the
observed spectrum and a power-law continuum (top curve), Fe
II-subtracted spectrum (middle curve), and Fe II spectrum (buttom
curve).}
\end{center}
\end{figure}

\section{RESULTS}
\begin{figure}
\begin{center}
\includegraphics[width=5cm,angle=-90]{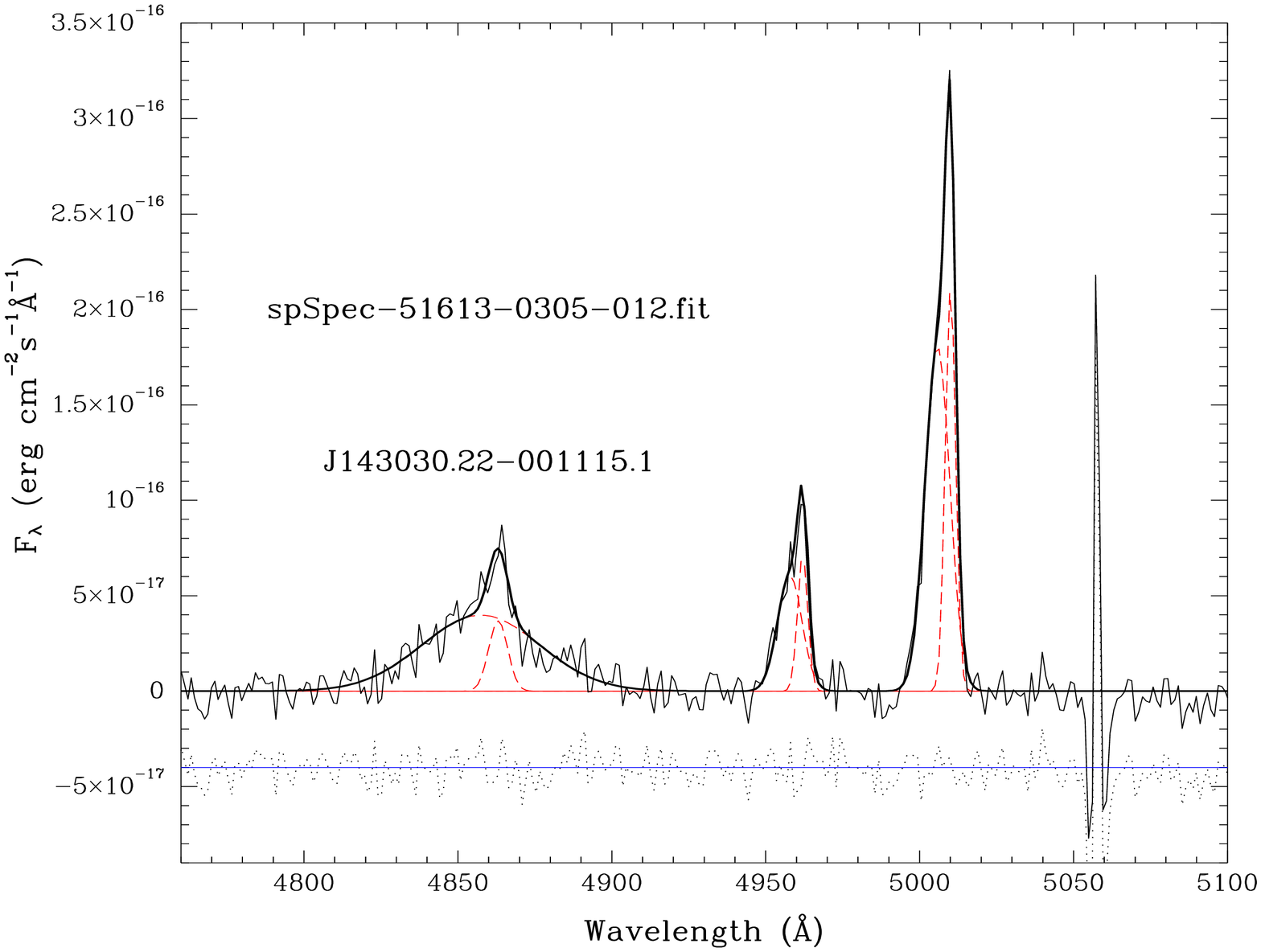}
\includegraphics[width=5cm,angle=-90]{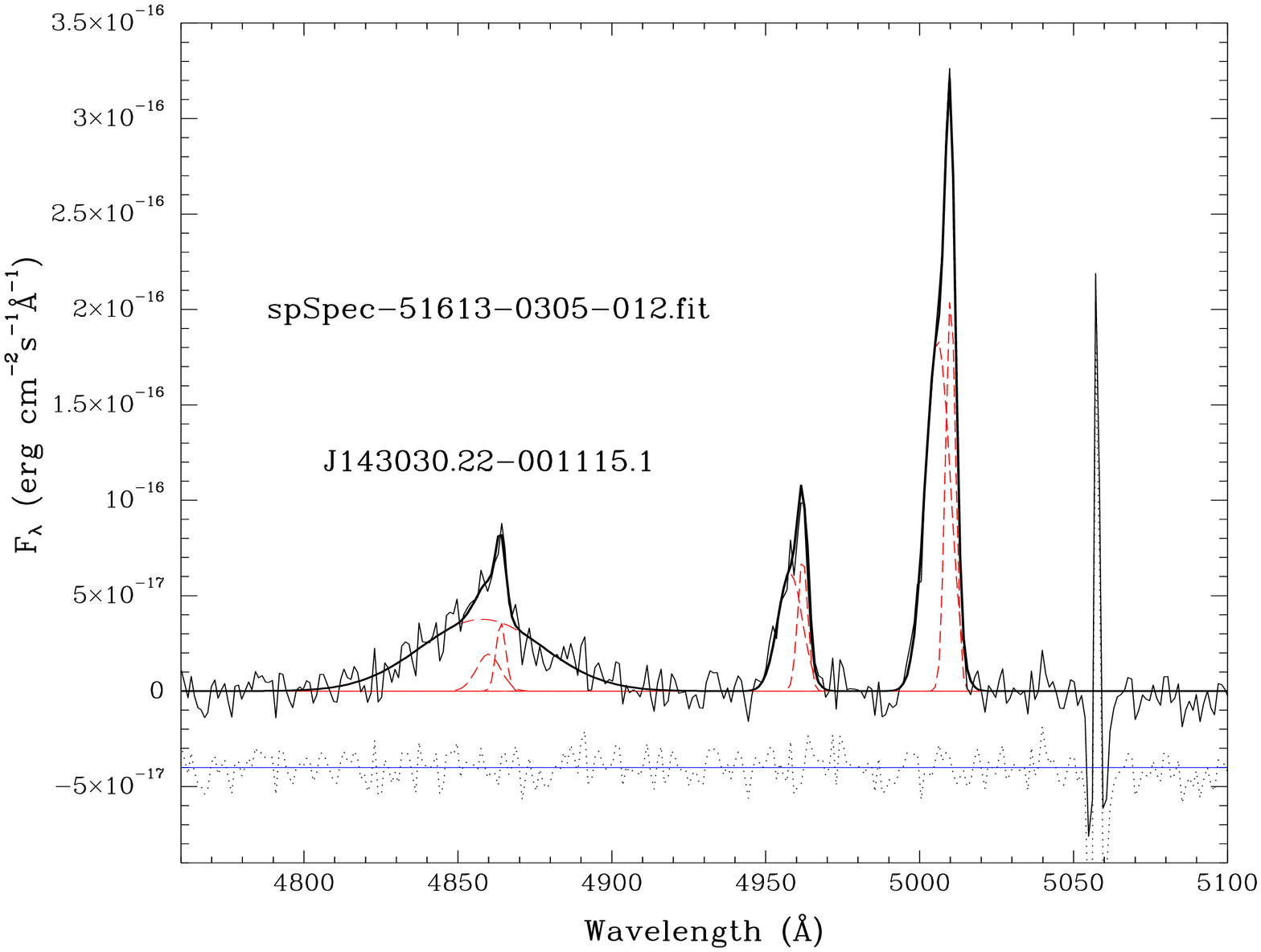}

\label{fig2} \caption{Multi-component fitting of the H$\beta$ and
[O III]$\lambda \lambda$ 4959, 5007: modelled composite profile
(thick solid line), individual components (the dotted lines), the
residual spectrum (lower panel). Left: two components to model
the H$\beta$ line; Right: three components to model the H$\beta$
line.}
\end{center}
\end{figure}

For SDSS J143030.22-001115.1, the [O III]$\lambda$5007 is very
strong relative to $H\beta$ and the Fe II multiples are not too
strong (See Fig. 1). The flux ratio of Fe II (between 4434 and
4684\AA) to total H$\beta$ (Fe II$\lambda$ 4570/H$\beta$) is
0.59$\pm$ 0.17, which is much smaller than the mean value for the
sample of 150 NLS1s (See Fig. 2 in Bian, Yuan \& Zhao 2005).

In the left panel of Fig. 2, we showed two-component fitting of
H$\beta$ and [O III]$\lambda \lambda$ 4959, 5007 for the
rest-frame spectrum of SDSS J143030.22-001115.1. The FWHM, flux of
each components are listed in Table 1.

\begin{table*}
\begin{center}
\begin{tabular}{lcccccccc}
\hline
 \hline

Line & Component & Rest Wavelength & FWHM & Line flux \\
& &  ($\AA$) &(\kms) &($10^{-16} erg s^{-1} cm^{-2}$)\\
(1) & (2) &(3) & (4) & (5)  \\
\hline
\multicolumn{5}{c}{Two components to model the H$\beta$ line}\\
\hline

{H$\beta$}               & n& $4859.9\pm$ 1.1    &$447\pm 199$& $2.8\pm1.1$ \\
                         & b& $4854.3\pm$ 1.1     &$2785\pm 276$& $19.1\pm1.2$ \\
{[O III]$\lambda$4959}   & n& $4960.6\pm$ 0.1     &$238\pm 21$&  $3.0\pm 0.5$ \\
                         & b& $4956.5\pm$ 0.4     &$563\pm 27$&  $6.0\pm 0.5$ \\
{[O III]$\lambda$5007}   & n& $5006.8\pm$ 0.1      &$238\pm 21$& $9.0\pm 1.5$ \\
                         & b& $5002.60\pm$ 0.4     &$563\pm 27$& $18.0\pm 1.5$ \\
\hline
\multicolumn{5}{c}{Three components to model the H$\beta$ line}\\
\hline

{H$\beta$}               & n1& $4862.6\pm$ 0.1    &$231\pm 21$& $1.4\pm0.3$ \\
                         & n2& $4858.6\pm$ 0.4     &$567\pm 25$& $1.9\pm0.6$ \\
                         & b& $4856.4\pm$ 1.2      &$2911\pm 210$& $18.9\pm1.1$ \\
{[O III]$\lambda$4959}   & n& $4960.6\pm$ 0.1      &$231\pm 21$&  $2.9\pm 0.4$ \\
                         & b& $4956.5\pm$ 0.4      &$567\pm 25$&  $6.2\pm 0.5$ \\
{[O III]$\lambda$5007}   & n& $5006.8\pm$ 0.1     &$231\pm 21$& $8.6\pm 1.2$ \\
                         & b& $5002.6\pm$ 0.4     &$567\pm 27$& $18.5\pm 1.5$ \\
\hline
\multicolumn{5}{c}{Starlight contribution and three components to model the H$\beta$ line}\\
\hline

{H$\beta$}               & n1& $4860.8\pm$ 0.5    &$307\pm 44$& $2.4\pm2.6$ \\
                         & n2& $4855.4\pm$ 0.7     &$408\pm 56$& $1.3\pm2.7$ \\
                         & b& $4856.7\pm$ 1.4      &$2657\pm 154$& $15.7\pm3.3$ \\
{[O III]$\lambda$4959}   & n& $4958.8\pm$ 0.5      &$307\pm 44$&  $5.6\pm 1.9$ \\
                         & b& $4953.3\pm$ 0.7      &$408\pm 56$&  $4.3\pm 1.2$ \\
{[O III]$\lambda$5007}   & n& $5006.8\pm$ 0.5     &$307\pm 44$& $16.8\pm 5.6$ \\
                         & b& $5001.2\pm$ 0.7     &$408\pm 56$& $12.9\pm 3.7$ \\
\hline

\end{tabular}

\caption{Results of the multi-component profile fitting of SDSS
J143030.22-001115.1. Columns are: (1) emission line; (2) emission
line components, where ``n, n1, n2'' and ``b'' represent the
narrow and broad components, respectively; (3) the rest wavelength
of the line component in angstroms; (4) FWHM of the line
components (\kms); (5) integrated line flux ($10^{-16} erg s^{-1}
cm^{-2}$). }
\end{center}
\end{table*}

From left panel of Fig. 2 and Table 1, we found that the
line-intensity ratio of [O III]$\lambda$5007 to the narrow
H$\beta$ is 9.6$\pm$ 3.8, which is consistent with the universal
adopted value of 10 from the photoionization model for NLRs
clouds. Considering the errors of the linewidth, the FWHM of the
narrow H$\beta$ line is consistent with that of FWHM of the narrow
and broad [O III] lines. The FWHM of the broad H$\beta$ line is
2785$\pm$ 276 \kms.

The template built from [O III] or [S II] is usually used to model
narrow H$\beta$ and H$\alpha$ (Grupe et al. 1998; Grupe et al.
1999; Grupe et al. 2004; Greene \& Ho, 2005a; 2005b). Since we
used two-component profiles to model the asymmetric [O
III]$\lambda$5007 profile, we should use three-component profiles
to model H$\beta$, two components are coming from NLRs and one is
BLRs. We take the same linewidth for each component of H$\beta$
and [O III] emitted from NLRs. And the flux ratio of [O III] to
H$\beta$ from NLRs is set to be free. In the right panel of Fig.
2, we showed the three-component fitting of the H$\beta$ line. The
FWHM, flux of each components are also listed in Table 2. The
line-intensity ratio of [O III]$\lambda$5007 to the narrow
H$\beta$ line is 8.2$\pm$ 3.2. The FWHM of the broad H$\beta$ line
is 2911$\pm$ 210 \kms. Therefore, SDSS J143030.22-001115.1 is not
a genuine NLS1 when we subtracted the H$\beta$ contribution from
NLRs.

\section{DISCUSSION}

\subsection{Black hole mass correction}
NLS1s belong to seyfert 1 galaxies and were initially defined by
their optical characteristics: FWHM(H$\beta$) less than 2000 \kms,
which is usually used to calculate the central black hole virial
mass (e.g. Wang \& Lu 2001; Bian \& Zhao 2004). We should use the
line width of H$\beta$ coming from the BLRs to trace the BLRs
virial movement. For SDSS J143030.22-001115.1, Williams et al.
(2003) directly measured the width of the line halfway between the
fitted continuum and the H$\beta$ line peak and obtained
FWHM(H$\beta$) to be 1744 \kms. Considering the empirical
size-luminosity formula and FWHM(H$\beta$) (Kaspi et al. 2000), we
have calculate its central supermassive black hole mass using the
value of 1744 \kms (Bian \& Zhao 2004). Its mass is $10^{6.5}$
\Msolar and its Eddington ratio log$L_{bol}/L_{Edd}$ is -0.60.
Here we used the FWHM of broad H$\beta$ to recalculate the black
hole mass. Its mass would be 0.44 larger if we take 2911 \kms as
FWHM(H$\beta$), 0.41 larger if we take 2785 \kms as
FWHM(H$\beta$). And log$L_{bol}/L_{Edd}$ would be -1.04$\sim$
-1.01. If we used the FWHM of narrow [O III] line to trace the
bulge stellar velocity dispersion, we found SDSS
J143030.22-001115.1 follow the so-called $M_{bh}-\sigma$ relation,
$M_{\rm bh}=10^{8.13}(\frac{\sigma}{200 {\rm km~
s^{-1}}})^{4.02}\Msolar$ (Tremaine et al. 2002).

For seven NLS1s, Rodriguez-Ardila et al. (2000) found that the
narrow component of H$\beta$ (H$\beta_{n}$) is about, 50\% of the
total line flux and the [O III] $\lambda$5007/H$\beta_{n}$ ratio
emitted in the NLRs varies from 1 to 5, instead of the universally
adopted value of 10. We also found the [O III] is not too weak in
many SDSS NLSls. This is consistent with the results of a sample
of 64 NLSls presented by Veron-Cetty et al. (2001). When we
calculate the virial black hole masses of NLS1s, we should measure
the H$\beta$ linewidth after subtracting the H$\beta$ contribution
from NLRs, which is especially important for NLS1s with luminous
[O III] line. For the mass correction of other objects in the
sample of 150 NLS1s and their locus in $M-\sigma$ plane, please
referee to Bian, Yuan \& Zhao (2006). There is no consideration
about multi-component in $H\beta$ line in the traditional
definition of NLS1s by Osterbrock \& Pogge (1985) and Goodrich
(1998). Therefor, we should cations about this traditional
definition and just consider the H$\beta$ line from BLRs in this
definition to classify NLS1s.

\subsection{Starlight contribution}

The fibers in SDSS have a diameter of 3" on the sky, corresponding
to ~6.5kpc at the redshift of SDSS J143030.22-001115.1. Its SDSS
spectrum taken through such a fixed aperture includes most of the
emission from the host galaxy emission (petroRad$_r$=3".43). At
the same time, the nuclear luminosity is comparable with its host
galaxy (psfMag$_r$=18.85 and petroMag$_r$=17.9). Using the method
of Lu et al. (2005) (Li et al. 2005; Hao et al. 2005), the stellar
component is modelled. The results are showed in Fig. 3 and Table
3. The main conclusions are not changed.

\begin{figure}
\begin{center}
\includegraphics[width=5cm,angle=-90]{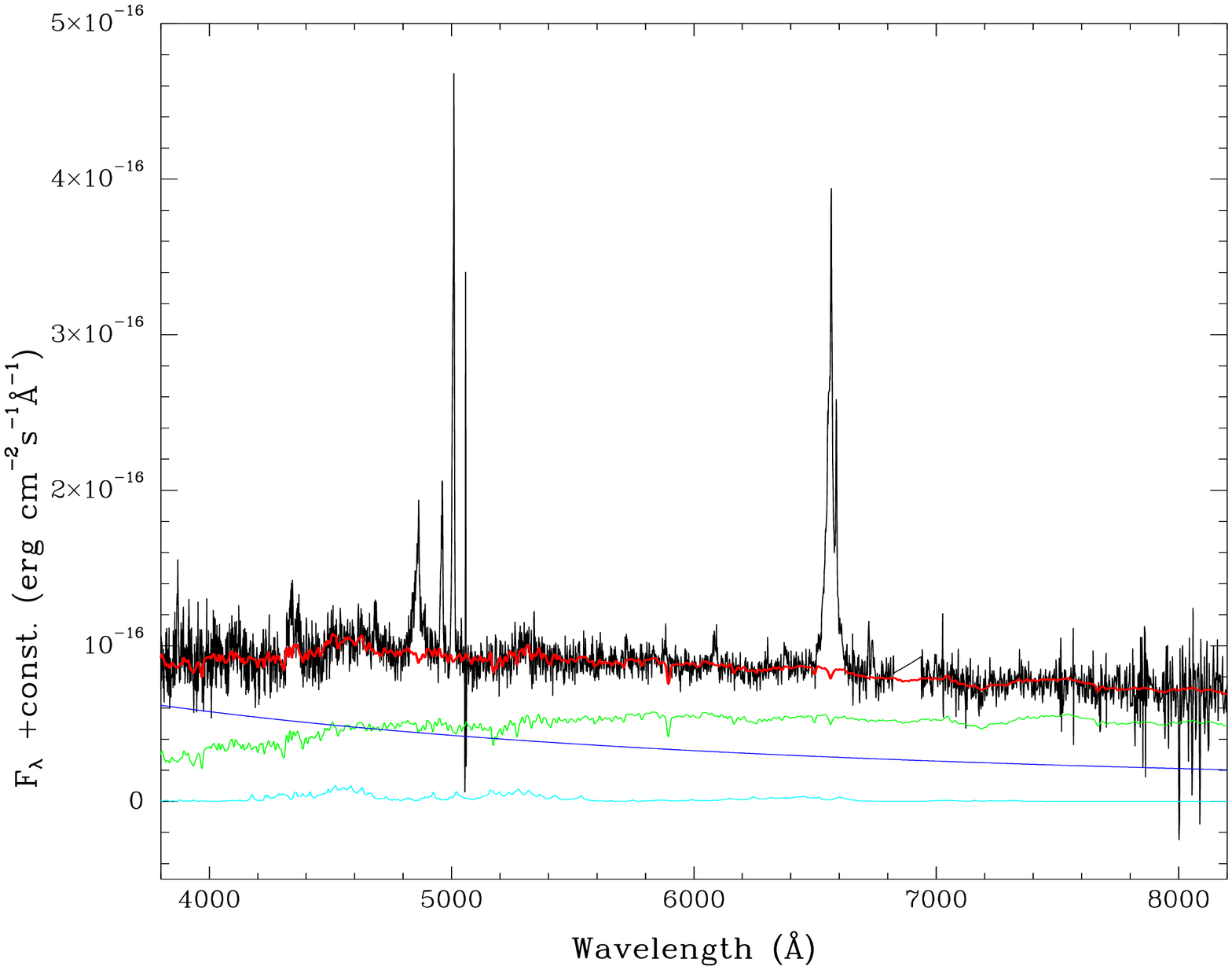}
\includegraphics[width=5cm,angle=-90]{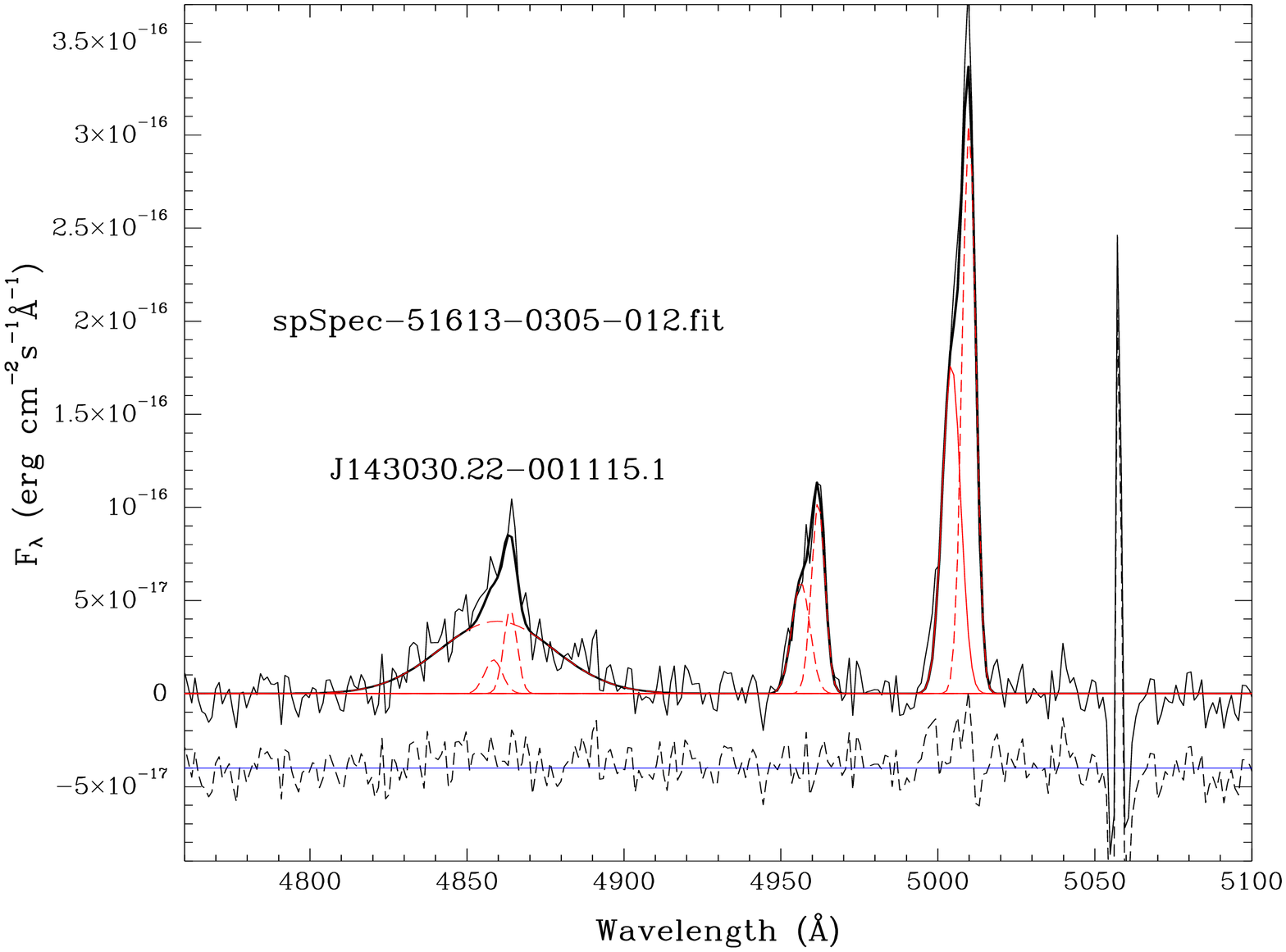}

\label{fig3} \caption{Left: the observed spectrum (black curve), a
power-law continuum (middle blue curve), Fe II spectrum (bottom
cyan curve), starlight contribution (middle Green curve), and
composition of continuum, Fe II and starlight (red curve). Right:
three components to model the H$\beta$ line.}
\end{center}
\end{figure}

\subsection{X-ray photon index}
ROSAT/ASCA X-ray observation of NLS1s showed that NLS1s generally
have softer X-ray spectra than other AGNs, i.e. the larger photon
index $\Gamma$, where $N_{E} \propto E^{-\Gamma}$ (e.g. Boller et
al. 1996; Leighly 1999). Williams et al. (2004) suggested it is
not always the case. They presented the Chandra observation of 17
NLS1s selected from the sample of 150 SDSS NLS1s (Williams et al.
2003). They derived $\Gamma$ from spectral fitting in Sherpa or
hardness ratio. They found two objects have $\Gamma$ less than
one: $0.25^{+0.80}_{-1.01}$ for SDSS J125943.59+010255.1, $0.92\pm
0.64$ for SDSS J143030.22-001115.1. For SDSS J125943.59+010255.1,
its net Chandra 0.5-8keV count rate is $0.003\pm0.001$ and the
uncertainties of $\Gamma$ is too large. For SDSS
J143030.22-001115.1, as we discussed above, it is not a genuine
NLS1. We should remove these two objects to do the statistics of
$\Gamma$ of NLS1s. The mean $\Gamma$ is $2.39 \pm0.15$ with the
standard deviation of 0.59 when these two objects are excluded.
Comparing with the result of Boller et al. (1996), the value of
$\Gamma$ is smaller, which is partially due to the the different
energy range coverage (Williams et al. 2004). It is suggested
there is a correlation between $\Gamma$ and the accretion
Eddington ratio (e.g. Bian \& Zhao 2003; Grupe 2004; Williams et
al. 2004; Bian 2005). There possibly exists NLS1s with smaller
Eddington ratio and flat X-ray spectra.

It has been found that the optical properties of some NLS1s with
flat X-ray spectrum are not quite different from those with steep
X-ray spectrum. The Eddington ratio of the former may be smaller
than the latter. NLS1s as presently defined are unlikely to be a
homogeneous class. The apparent scarcity of "NLS1s" with low
Eddington ratio should result in selection effect: in flux limited
surveys, it is difficult to find AGNs with small black hole and
low mass accretion rate.

\section*{ACKNOWLEDGMENTS}

We thank Luis C. Ho for his very helpful comments. We thank the
referee, Dr H. Y. Zhou, for his useful remarks. This work has been
supported by the NSFC (No. 10403005; No. 10473005; No. 10273007)
and NSF from Jiangsu Provincial Education Department (No.
03KJB160060). Funding for the creation and distribution of the
SDSS Archive has been provided by the Alfred P. Sloan Foundation,
the Participating Institutions, NASA, the National Science
Foundation, the US Department of Energy, the Japanese
Monbukagakusho, and the Max Planck Society. The SDSS Web site is
http:// www.sdss.org/. This research has made use of the NASA/IPAC
Extragalactic Database, which is operated by the Jet Propulsion
Laboratory at Caltech, under contract with NASA.

\end{document}